\documentclass{moriond}

\usepackage{subcaption}
\usepackage{amsmath}

\def\be{\begin{equation}}
\def\ee{\end{equation}}
\def\bea{\begin{eqnarray}}
\def\eea{\end{eqnarray}}

\def\tev{$\,$TeV}
\def\gev{$\,$GeV}
\def\mev{$\,$MeV}
\def\ifb{$\,$fb${}^{-1}$}
\def\syst{\text{syst}}
\def\stat{\text{stat}}
\def\BF{\text{BF}}

\newcommand{\Photo}{}

\begin{document}
\vspace*{4cm}
\title{Rare Beauty and Charm Decays}

\author{ Angel Campoverde on Behalf of the LHCb Collaboration }

\address{University of the Chinese Academy of Sciences,\\
19A Yuquan Road, Beijing, China}

\maketitle\abstracts{
    There are two main ways of looking for new physics, direct searches and precision measurements.
    The latter are sensitive to a broader spectrum of models; they also can be sensitive to higher 
    energy scales than what can be reached through direct searches. We will provide an update of 
    several precision measurements carried out by the LHCb collaboration. In particular, we will show the
    measurement of $R_{\phi\pi}^{(s)}$, the search for $B_c^+ \to \pi^+\mu^+\mu^-$ and $B_s\to\mu^+\mu^-\gamma$ 
    and an amplitude analysis of the $\Lambda_b^0\to pK^-\gamma$ decay.
}

\section{Introduction}

LHCb is one of the experiments at CERN and its main goals include the study of beauty and charm decays.
The measurements in the following sections use proton-proton collision data collected from 2011 to 2018 at 
7\tev{}, 8\tev{} and 13\tev{}, amounting in total to 9\ifb{}. Although not every analysis uses the whole dataset. 

Rare decays are a powerful tool that can be used to search for new physics. They benefit from both
a low Standard Model background and from the partial cancellation of systematic effects, due to the fact that
the observables are measured as ratios of correlated quantities. On the other hand, an anomaly's
interpretation as new physics is not as straightforward as the presence of an excess in a mass spectrum would 
be in a direct search. Despite the partial cancellation of systematic effects, these measurements require a 
superb knowledge of detector effects. Extensive and detailed calibration studies need to be carried out in 
order to access these observables.

\section{Measurement of $\mathcal{B}(\phi\to\mu^+\mu^-)/\mathcal{B}(\phi\to e^+e^-)$}
This analysis measures the ratio of branching ratios for $D_{(s)}^+\to\pi^+\phi(\ell^+\ell^-)$ decays between 
the final states with muons and electrons, normalized by the $B^+\to K^+J/\psi(\to \ell^+\ell^-)$ decay:

\begin{equation*}
    R_{\phi\pi}^{(s)}=\frac{\mathcal{B}(D_{(s)}^+\to\pi^+\phi(\mu^+\mu^-))}{\mathcal{B}(D_{(s)}^+\to\pi^+\phi(e^+e^-))} \Bigg/ 
    \frac{\mathcal{B}(B^+\to K^+J/\psi(\to \mu^+\mu^-))}{\mathcal{B}(B^+\to K^+J/\psi(\to e^+e^-))}.
\end{equation*}

Given that 
$\phi(1020)\to\ell^+\ell^-$ and 
$J/\psi\to\ell^+\ell^-$ 
are dominated by the electromagnetic interaction, it is expected that both the branching fraction ratio involving 
$D_{(s)}^+$ and the one involving $B^+$ decays be unity. The $B^+$ decays are also used to extract corrections 
to simulation that will be applied to the $D_{(s)}^+$ decays. This study is aimed at confirming the 
portability of the corrections from the $q^2$ bin
\footnote{$q^2$ will be defined, from now on, as the square of the mass of the dilepton.}
associated with the $J/\psi$ meson to the one associated with
the $\phi(1020)$ meson. Thus, the usage of the $B^+$ decay for normalization purposes would serve to reduce 
biases in the calibration.

Given the soft nature of the leptons in this study,
the trigger efficiency would be too low if the analysis would target events triggered by only the signal leptons. Therefore, also events
triggered by the pion and any object not associated with the signal candidate are used.
The measurement of the branching fractions requires extracting the yields 
through fits to the $D_{(s)}^+$-candidate invariant mass distribution.
The leptons in this analysis originate from an intermediate resonance, 
the $\phi(1020)$ meson. Therefore, it is possible to 
use a mass constraint around its PDG\cite{pdg} value to improve the resolution of the $D_{(s)}^+$ candidate's mass.

\subsection{Background estimation}
The main backgrounds are $D^+\to K^+_{\to e^+}\pi^-_{\to e^-}\pi^+$ and $D^+\to \pi^+_{\to e^+}\pi^-_{\to e^-}\pi^+$ decays
\footnote{Here the subscript $x_{\to e}$ symbolizes a particle $x$ that has been reconstructed as an electron.}.
In the former case, the background is reduced by assigning to the electrons the $K^+$ and $\pi^-$ mass hypotheses and then applying 
a veto around the mass of the $D^+$. In the latter case, the reduction happens by using particle identification (PID) requirements; 
however this background cannot be ignored in the fit, thus one needs to model it. For that reason, a control region is built by 
inverting the PID requirements on the electrons, as shown in Fig~\ref{fig:mid_cr}. From this control region, the shape of the 
background is extrapolated to the signal region.
Additionally, the combinatorial background is modelled in the fit for both the electron and the muon channels. In the electron channel, the 
mass constraint on the dielectron is responsible for a warping of the normally exponential shape, thus a third degree polynomial is 
used to fit this component.

\begin{figure}[ht]
    \centering
    \begin{subfigure}{0.42\textwidth}
	\includegraphics[width=\linewidth]{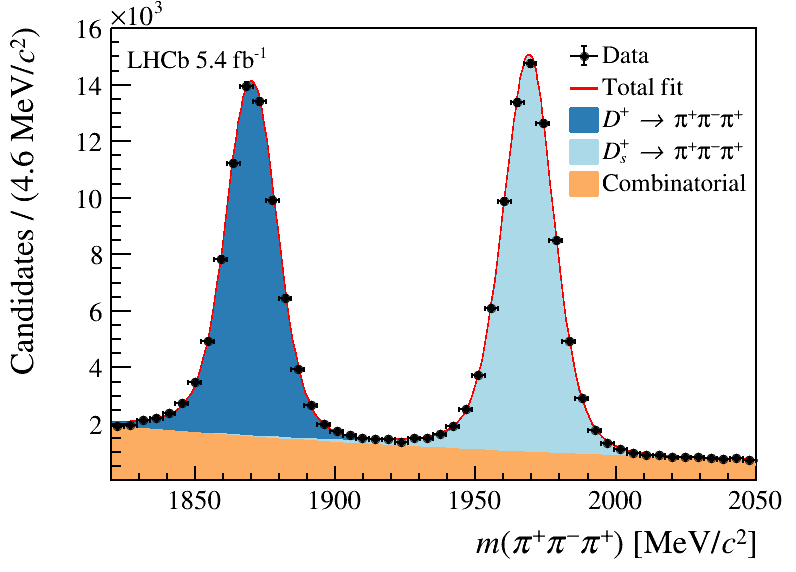}
	\caption{}
	\label{fig:mid_cr}
    \end{subfigure}
    \begin{subfigure}{0.42\textwidth}
	\includegraphics[width=\linewidth]{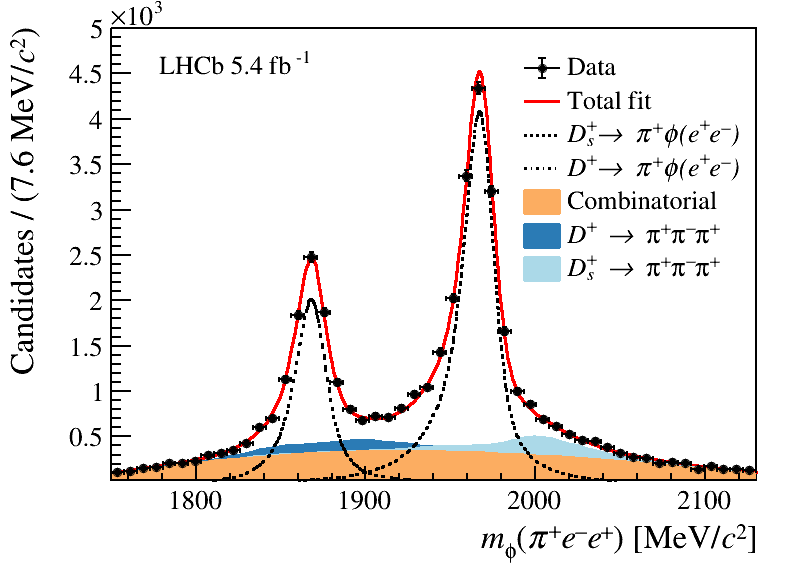}
	\caption{}
	\label{fig:sig_ee}
    \end{subfigure}
    \caption{
	On the left, fit to misID backgrounds in a PID inverted control region.
    On the right, fit to the signal region in the electron channel.}
\end{figure}

\subsection{Systematics and results}
The fit to the signal region in the electron channel can be seen in figure~\ref{fig:sig_ee}.
The main systematics in this analysis stem from the simulation's modelling of the detector 
occupancy and the $q^2$ resolution. Both of these quantities are studied with simulation, corrected using 
collision data from the normalization mode. However, these corrections do not allow a full cancellation 
of the systematics, given the difference in the kinematics of the signal and normalization modes.

The values of $R_{\phi\pi}$, measured separately with $D^+$ and $D_s^+$ decays are:

\begin{equation*}
    R_{\phi\pi}^d=1.026\pm0.020~(\stat)\pm0.056~(\syst),
\end{equation*}
\begin{equation*}
    R_{\phi\pi}^s=1.017\pm0.013~(\stat)\pm0.051~(\syst).
\end{equation*}

While their combination is:

\begin{equation*}
    R_{\phi\pi}=1.022\pm0.012~(\stat)\pm0.048~(\syst).
\end{equation*}

From this value, together with the measured branching fraction of $\phi\to e^+e^-$~\cite{pdg}, the most precise 
measurement of the branching fraction of $\phi\to\mu^+\mu^-$ is obtained as:
\begin{equation*}
\mathcal{B}(\phi\to\mu^+\mu^-)=(3.045\pm0.049~(\stat)\pm0.148~(\syst))\times 10^{-4}.
\end{equation*}

\section{Measurement of $R_{\pi^+\mu^+\mu^-/J/\psi}$ and $R_{\psi(2S)/J/\psi}$}
In this section we will present the first search for the non resonant decay $B_c^+\to \pi^+\mu^+\mu^-$. This search was carried out using
the full 9\ifb{} of data collected by the LHCb detector. The events collected are required to have a signal muon firing the trigger. 
As done in the past section, this measurement profits from the use of a 
normalization decay, $B_c^+\to \pi^+ J/\psi(\to \mu^+\mu^-)$, to partially cancel systematic effects. Thus, the quantity measured can be
written as:

\begin{equation*}
    R_{\pi^+\mu^+\mu^-/J/\psi}\equiv\frac{\mathcal{B}(B_c^+\to\pi^+\mu^+\mu^-)}{\mathcal{B}(B_c^+\to J/\psi(\to\mu^+\mu^-)\pi^+)}.
\end{equation*}

As a validation of the efficiencies, the normalization mode is also used to measure:

\begin{equation*}
    R_{\psi(2S)/J/\psi}\equiv\frac{\mathcal{B}(B_c^+\to\psi(2S)\pi^+)}{\mathcal{B}(B_c^+\to J/\psi\pi^+)}.
\end{equation*}

The analysis is done in four bins of $q^2$ and a mass constrain on the dimuon system around the charmonium's PDG mass is applied 
to measure $R_{\psi(2S)/J/\psi}$. Furthermore a mass window of 50\mev{}$/c^2$ around the charmonium is used to select the resonant modes.

\subsection{Background estimation and selection}

In the fit to the resonant modes, the partially reconstructed components 
$B_c^+\to\rho^+\mu^+\mu^-$, $B_c^+\to J/\psi \rho^+$ and $B_c^+\to\psi(2S)\rho^+$ with $\rho\to\pi^+\pi^0$
are included. Potential backgrounds, where double misidentification happens, like 
$B_c^+\to\pi^+\pi^-\pi^+$ or $B_c^+\to c\bar{c}(\to \mu^+_{\to \pi^+}, \mu^-) \pi^+_{\to\mu^+}$ 
are suppressed by PID requirements. In the signal region fit, only the 
combinatorial background is modeled, the other backgrounds are negligible.

The selection uses a BDT trained with simulated signal events and data in the sidebands of 
the signal region as proxies for the signal and background respectively. 

\subsection{Results}
Fig.~\ref{fig:rpmm_fit} shows the fit to the signal region and the normalization dataset. As can be seen, 
no signal is present and therefore upper limits are set with the Feldman-Cousins approach.

\begin{figure}[ht]
    \centering
    \begin{subfigure}{0.42\textwidth}
        \includegraphics[width=\linewidth]{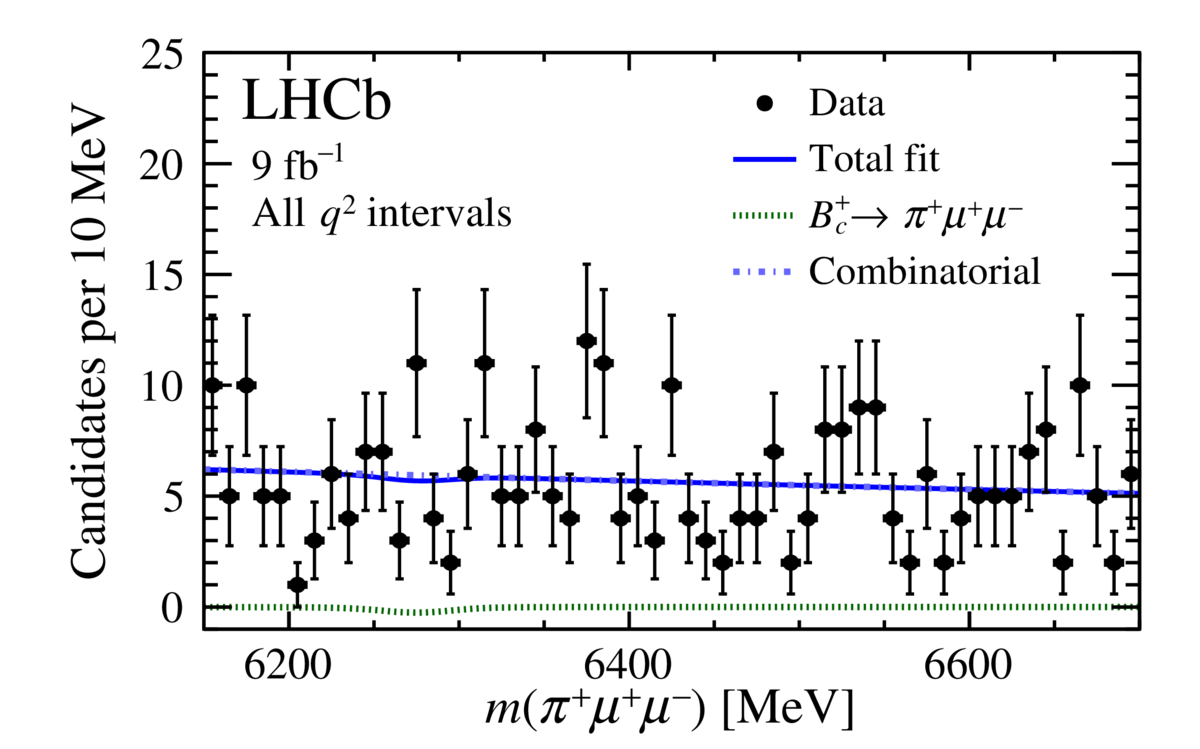}
    \end{subfigure}
    \begin{subfigure}{0.42\textwidth}
        \includegraphics[width=\linewidth]{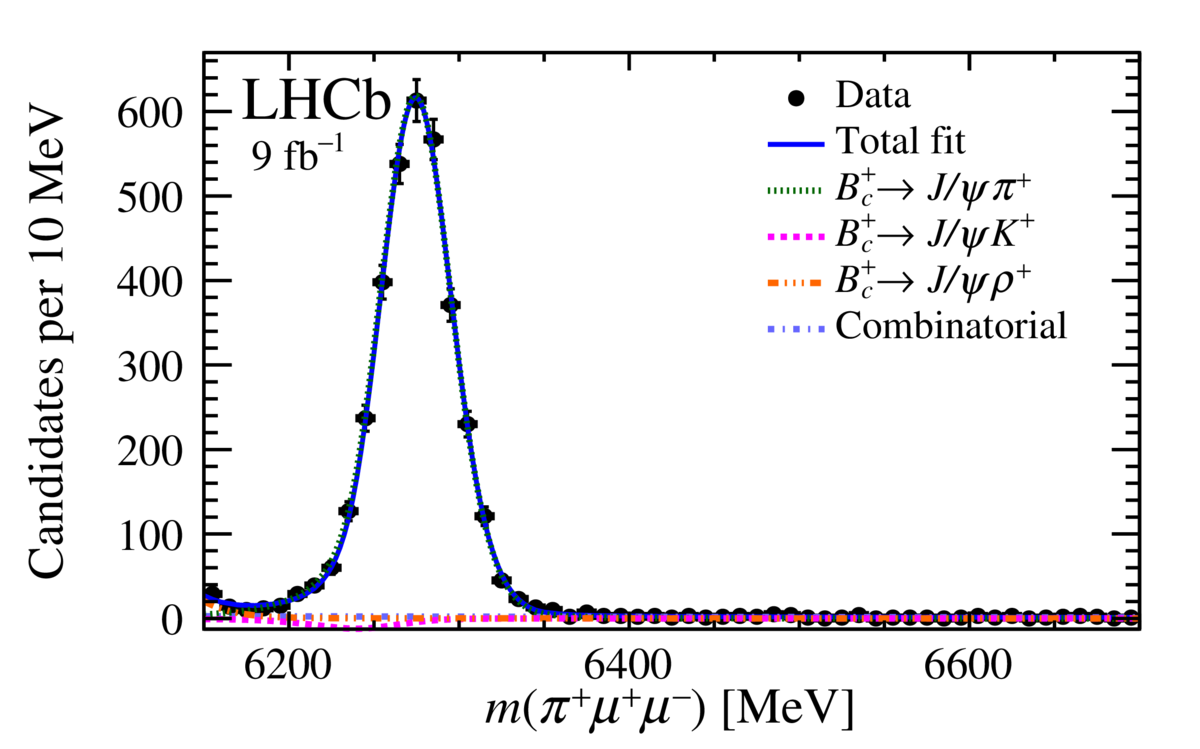}
    \end{subfigure}
    \caption{Fits to the signal (left) and normalization (right) datasets.}
    \label{fig:rpmm_fit}
\end{figure}

There is no theoretical prediction for the branching fraction of $B_c^+\to\pi^+\mu^+\mu^-$, and
the result of this analysis:

\begin{equation*}
    \frac{\mathcal{B}(B_c^+\to\pi^+\mu^+\mu^-)}{\mathcal{B}(B_c^+\to\pi^+J/\psi)}<2.1\times10^{-4},
\end{equation*}

is the first upper limit for this ratio. While:

\begin{equation*}
    \frac{\mathcal{B}(B_c^+\to\pi^+\psi(2S))}{\mathcal{B}(B_c^+\to\pi^+J/\psi)}=0.254\pm0.018~(\stat)\pm0.03~(\syst)\pm0.005~(\BF),
\end{equation*}

is the most precise measurement to date.

\section{Search for the $B_s^0\to \mu^+\mu^-\gamma$ decay}
This search uses 5.4\ifb{} of data collected at $\sqrt{s}=13$\tev{}. The measurement uses events in which 
either the candidate's decay products or particles not associated with the candidate fire the trigger.
Unlike in $B_s^0\to\mu^+\mu^-$, the presence of the photon allows the dimuon invariant mass to go lower; this in turn allows to explore
the operators $\mathcal{O}_7^{(')}$, $\mathcal{O}_{9,10}^{(')}$, $\mathcal{O}_{1,2}$ and $\mathcal{O}_{9,10}^{(')}$. 
The regions where the analysis is sensitive to different operators are shown in~Fig.\ref{fig:bsmmg_q2}. The presence
of the photon also lifts the chiral suppression present in $B_s^0\to\mu^+\mu^-$, such that both decays have branching fractions at the
same order of magnitude.

Upper limits of: 
\begin{equation*}
    \mathcal{B}(B_s^{0}\to\mu^+\mu^-\gamma) < 2\cdot 10^{-9},
\end{equation*}
at 95\% confidence level have been already set by~\cite{bsmm} for which this process is a partially reconstructed background and 
where the photon was not reconstructed; also it is limited to $m(\mu^+,\mu^-)>4.9\,\text{GeV}/c^2$. However this is the first dedicated search for this decay. As was done in the last section, this
analysis also carries out the search in bins of $q^2$ as shown in table~\ref{tab:binning}.
The results are obtained with and without a veto around the $\phi(1020)$ mass. 

\begin{table}[tbh]
    \caption{\small Mass range definition, predicted branching fraction, and the expected fraction of signal yield in the different $q^2$ bins.} 
    \centering \small
    \begin{tabular}{lccc}
	\hline\hline
	\rule{0pt}{2.5ex}$q^2$ bin~& I & II & III \\ 
	\hline 
	\rule{0pt}{2.5ex}$q^2$ [\gev{}${}^2/c^{2}$]& $[4m_{\mu}^2, 2.89]$ & $[2.89,8.29]$ & $[15.37,m_{B_s}^2]$ \\
	$m(\mu^+, \mu^-)$ [\gev{}${}/c^2$] & $[2m_{\mu},1.70]$ & $[1.70,2.88]$ & $[3.92,m_{B_s}]$ \\
	$10^{10}\times\mathcal{B}(B_s\to\mu^+\mu^-\gamma)$& $82\pm15$ & $2.54\pm0.34$ & $9.1\pm1.1$\\
	Fraction of $B_s\to\mu^+\mu^-\gamma$ & 87\% & 2.7\% & 9.8\%\\
	\hline\hline
    \end{tabular}
    \label{tab:binning}
\end{table}

The analysis uses $B_s^0\to J/\psi(\to\mu^+\mu^-)\eta(\to \gamma\gamma)$ as the normalization decay due to its
well known branching fraction and its similarity to the signal decay. While the control mode was chosen as 
$B_s^0\to\phi(\to K^+K^-)\gamma$ due to its large statistics and its three body topology, also present 
in the signal. 

\subsection{Background estimation and selection}
Both photons and muons have to satisfy minimum $p_T$ requirements, while the muons are also required to satisfy
PID criteria. Requirements are also imposed on the vertex fit quality and the $p_T$ of the $B_s^0$. 

This study includes in the fitting model combinatorial, partially reconstructed 
and peaking backgrounds. The former is modelled with an exponential
and corresponds to candidates made from random combinations of photons and muons. The latter is made of 
decays where at least one particle was not reconstructed. These candidates show up as a broad distribution 
below the $B_s^0$ mass. Three partially reconstructed decays were considered 
$B_s^0\to\mu^+\mu^-\phi(\to\pi^+\pi^-\pi^0)$, 
$B_d\to\mu^+\mu^-\eta^{'}(\to \rho^0\gamma)$ and 
$B_d\to D^-\mu^+\nu_{\mu}$ with $D^-\to\pi^0\mu^-\nu_{\mu}$. 
The peaking backgrounds considered are 
$B_s^0\to\mu^+\mu^-\eta$,
$B^0\to\mu^+\mu^-\eta$ and
$B^0\to\mu^+\mu^-\pi^0$ decays.

\subsection{Results}
The fits to the signal region in the four $q^2$ bins show no evident signal, therefore upper limits are set
and these are shown in~Fig.\ref{fig:bsmmg_uplim}. The values with and without the $\phi(1020)$ veto on the 
lowest $q^2$ bin at 90\% (95\%) CL are respectively:

\begin{equation*}
    \mathcal{B}(B_s^0\to\mu^+\mu^-\gamma)<2.9(3.4)\times10^{-8},
\end{equation*}

\begin{equation*}
    \mathcal{B}(B_s^0\to\mu^+\mu^-\gamma)<2.5(2.8)\times10^{-8}.
\end{equation*}

\begin{figure}[ht!]
    \centering
    \begin{subfigure}{0.49\textwidth}
	\includegraphics[width=\linewidth]{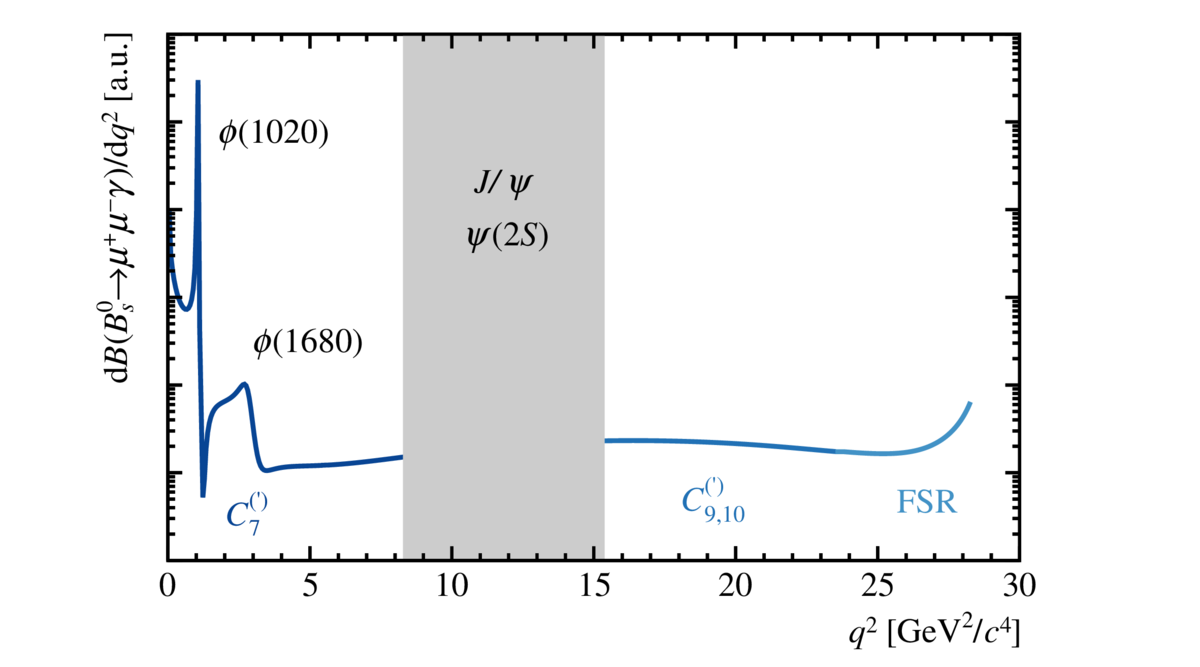}
	\caption{}
	\label{fig:bsmmg_q2}
    \end{subfigure}
    \begin{subfigure}{0.41\textwidth}
	\includegraphics[width=\linewidth]{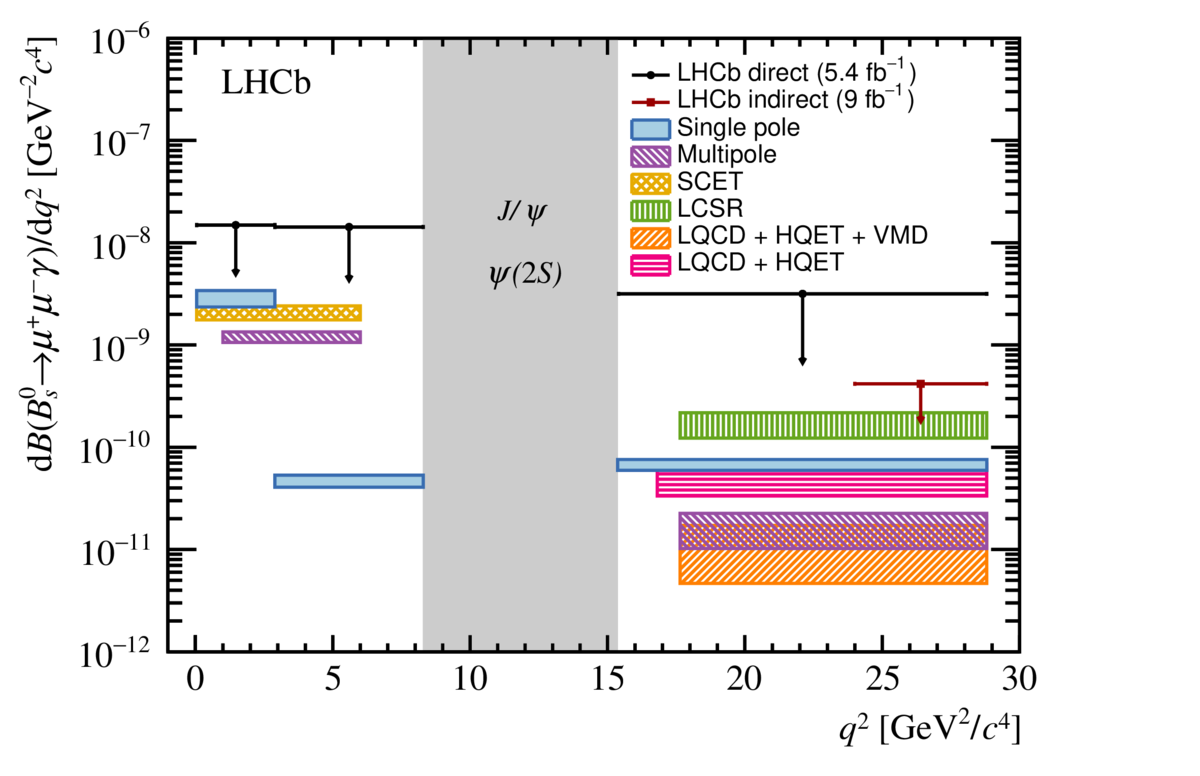}
	\caption{}
	\label{fig:bsmmg_uplim}
    \end{subfigure}
    \caption{
	On the left, $q^2$ spectrum alongside the operators to which the analysis is sensitive in each region.
	On the right, upper limits on the branching ratio for $B_s^{0}\to\mu^+\mu^-\gamma$ and theoretical predictions.
    }
\end{figure}

\section{Amplitude analysis of the $\Lambda_b^0\to pK^-\gamma$ decay}
This analysis uses the full 9\ifb{} of data collected by LHCb. A similar study has been carried out 
with $\Lambda_b^0\to pK^- J/\psi$~\cite{pent}.
However the presence of a photon allows for the $pK$ system mass to reach values of up to $2.5$\gev{}$/c^2$. 
The analysis strategy consists on performing an unbinned maximum likelihood fit to the 
reconstructed $\Lambda_b^0$ invariant mass spectrum
in fully selected data to separate signal from background and then perform background subtraction 
with the sPlot~\cite{splt} technique. This background subtracted data is later used in the amplitude analysis.

\subsection{Background estimation and selection}
The mass fit models combinatorial and partially reconstructed backgrounds. 
Both $D^0\to KK$ and $D^0\to K\pi$ can contribute to the background when combined with a random photon 
and when one of the hadrons is misidentified. However vetoing around the $D^0$ mass distorts the acceptance and thus 
these decays have been modeled in the fit. Another decay included in the fit is
the partially reconstructed $\Lambda_b^0\to pK^{*-}(\to K^-\pi^0)\gamma$.
The fits are performed separately for Run I and Run II data, due to the large differences in the data taking conditions.

\subsection{Results}
The model used in the amplitude analysis was taken from~\cite{ammod}.
Only contributions from well established $\Lambda$ resonances are taken into account, alongside a non-resonant component
which was seen to improve the quality of the fit.
The amplitude analysis uses a 2D unbinned maximum likelihood fit with observables as $m_{\Lambda_b^0}(pK)$ and 
$m_{\Lambda_b^0}(p\gamma)$ masses. The subscript here means that the mass of the $\Lambda_b^0$ candidate was constrained to the 
PDG mass in order to improve the resolution. The means and widths of the resonances were taken as external inputs
and these constitute the leading source of uncertainty in the analysis as can be seen in Fig.~\ref{fig:amfrac}. While the subleading
uncertainties stem from sources such as the amplitude model, the uncertainty in the acceptance or the limited size of the simulated 
samples. Multiple local minima are present, with different values for the nuisance parameters; however the fit fractions and 
interference amplitudes show consistent values.

\begin{figure}[ht]
    \centering
    \includegraphics[width=0.55\linewidth]{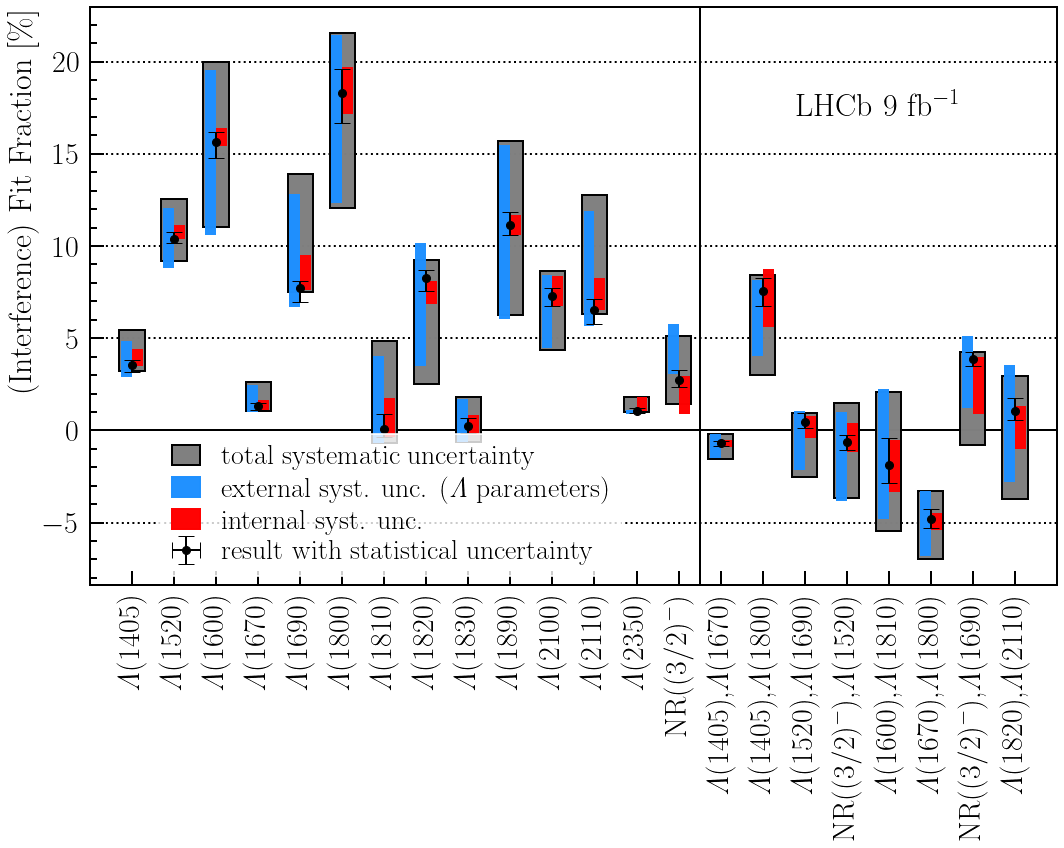}
    \caption{Decay amplitudes for transitions through several intermediate states in $\Lambda_b^0\to pK^-\gamma$ decays
    as well as the interferences.}
    \label{fig:amfrac}
\end{figure}

\section{Summary}
We have shown the most precise measurement of the $\phi(1020)\to\mu^+\mu^-$ branching fraction, the corresponding study also  
proved that efficiencies at low $q^2$ are well understood by the LHCb collaboration. We have also shown the
first upper limits on the branching fraction of $B_c^+\to\pi^+\mu^+\mu^-$ decays. The first dedicated search for
$B_s^0\to\mu^+\mu^-\gamma$ was also discussed, which extends previous studies of the $B_s^0\to\mu^+\mu^-$ process. 
Finally, a thorough amplitude analysis of $\Lambda_b^0\to pK^-\gamma$ decays has provided new insights into the contributions
of different $\Lambda$ baryons to this process. These studies will greatly benefit from the data that LHCb will collect in Run 3.


\section*{References}

\begin{thebibliography}{99}
    \bibitem{rphipi} LHCb Collaboration, submitted to JHEP, \color{blue}[arXiv:2402.01336]\color{black}.
    \bibitem{rpimm}  LHCb Collaboration, submitted to JHEP, \color{blue}[arXiv:2312.12228]\color{black}.
    \bibitem{bsmmg}  LHCb Collaboration, submitted to JHEP, \color{blue}[arXiv:2404.03375]\color{black}.
    \bibitem{aman}   LHCb Collaboration, submitted to JHEP, \color{blue}[arXiv:2403.03710]\color{black}.
    \bibitem{pent}   LHCb Collaboration, Phys Rev Lett, 115, 072001 (2015). 
    \bibitem{rk}     LHCb Collaboration, Phys Rev Lett, 131, 051803 (2023).
    \bibitem{bsmm}   LHCb Collaboration, Phys Rev D, 105, 012010 (2022).
    \bibitem{ammod} Albrecht, J., Amhis, Y., Beck, A. et al., JHEP, 06, 116 (2020).
    \bibitem{splt}  Muriel Pivk, Francois R. Le Diberder, Nucl.Instrum.Meth.A555:356-369 (2005)
    \bibitem{pdg}   P.A. Zyla et al. (Particle Data Group), Prog. Theor. Exp. Phys. 2020, 083C01 (2020) 
\end{thebibliography}
\end{document}

\typeout{get arXiv to do 4 passes: Label(s) may have changed. Rerun}


